# Toward a fractal spectrum approach for neutron and gamma pulse shape discrimination

Mingzhe Liu[1,2]*, Bingqi Liu[2], Zhuo Zuo[2], Lei Wang[2], Guibin Zan[2], Xianguo Tuo[1]*

[1] *State Key Laboratory of Geohazard Prevention and Geoenvironment Protection (Chengdu University of Technology), Chengdu 610059, China*

[2] *College of Nuclear Technology and Automation Engineering, Chengdu University of Technology, Chengdu 610059, China*

**Abstract** Accurately selecting neutron signals and discriminating gamma signals from a mixed radiation field is a key research issue in neutron detection. This paper proposes a fractal spectrum discrimination approach by means of different spectral characteristics of neutrons and gamma rays. Figure of merit and average discriminant error ratio are used together to evaluate the discrimination effects. Different neutron and gamma signals with various noise and pulse pile-up are simulated according to real data in the literature. The proposed approach is compared with the digital charge integration and pulse gradient methods. It is found that the fractal approach exhibits the best discrimination performance, followed by the digital charge integration method and the pulse gradient method, respectively. The fractal spectrum approach is not sensitive to high frequency noise and pulse pile-up. This means that the proposed approach has superior performance for effective and efficient anti-noise and high discrimination in neutron detection.

## I. INTRODUCTION

Neutron detection has attracted much attention in the areas of homeland security, environmental monitoring, metallurgy and building materials, radiation medicine, archaeology, agriculture and nuclear power [1]. A key research issue in neutron detection is how to quickly and accurately select out neutron signals and discriminate gamma signals from a mixed radiation field. Naturally, gamma-rays and neutrons are generated simultaneously, and electrically neutral neutrons interact with surrounding substances by inelastic scattering and radiation capture by moderated neutrons, thus in almost all situations, neutrons will be accompanied by a large gamma background. Neutron-sensitive detectors are also sensitive to gamma-rays; as a result it is difficult to distinguish signals from neutrons and gamma-rays [2-5].

On the basis of liquid scintillation detectors, many scholars have developed traditional analog circuits such as the zero-crossing method，time rise method and digital charge integration method to attempt to discriminate neutrons and gamma-rays. In recent years, with the development of Digital Signal Processing (DSP) and Field Programmable Gate Arrays (FPGAs) and the wide application of artificial intelligence, digitalized discrimination methods for neutrons and gamma-rays, such as the pulse gradient method, neural network method and wavelet analysis method have been developed [6-22]. These methods have led to a great improvement in accuracy in discrimination of neutrons and gamma-rays. This issue still deserves further study, and development of new algorithms containing more physical significance is necessary.

---

*Supported by the National Natural Science Foundation of China (41274109), Sichuan Youth Science and Technology Innovation

Research Team (2015TD0020), Scientific and Technological Support Program of Sichuan Province (2013FZ0022), and the Creative Team Program of Chengdu University of Technology.
*Corresponding author, E-mail: *liumz@cdut.edu.cn; txg@swust.edu.cn*.

Inspired by the different spectral characteristics of neutrons and gamma-rays, an n/γ discrimination approach based on a fractal spectrum is presented. The proposed approach is compared with the digital charge integration method and pulse gradient method in two aspects: figure of merit (FoM) and average discriminant error ratio (DER). Simulation data are used based on a specific detector, EJ-301 scintillator. It is found that the fractal approach exhibits better discrimination performance than the digital charge integration method and the pulse gradient method. In addition, this paper also discusses the influence of high frequency noise and pulse pile-ups on the three methods. This approach has the advantages of anti-noise and high discriminating ability.

## II. FRACTAL SPECTRUM METHOD

Fractal theory has been applied to many scientific fields including nuclear science and technology such as neutron transport [23], containment modeling [24] and heat transfer modeling in a nuclear reactor [25]. This theory extends integer discrete dimensions which contain points, lines, surfaces, and volumes in traditional Euclidean geometry into continuous non-integer dimensions. It can thus better display the irregularities and complexity of the real world by definition of a fractal dimension *D*. Fractal dimension is a very significant parameter which reflects the effectiveness of complex geometry space [23-26]. In fractals, scale invariance in the frequency domain manifests as an unchanging shape of the frequency domain. The spectral characteristics and cut-off frequency follow an exponential law. In general, the expression can be written as follows:

$$P(\omega) = C\omega^{(aD-b)} \qquad (1)$$

where $\omega$ is the frequency, *C* is the scale factor, *D* is the fractal dimension, and *a* and *b* are undetermined coefficients. Note that $a = 2$ and $b = 5$ are specified in Ref. [26]. Using the Fourier transform to calculate the corresponding frequency, one finally gets the spectrum of $P(\omega)$. Its fast Fourier transform (FFT) expression can be written as:

$$X(k) = \sum_{j=1}^{N} x(j)\omega_N^{(j-1)(k-1)} \qquad (2)$$

where $X(k)$ is the spectral data of after FFT, $x(j)$ is the spectrum data before FFT, $\omega_N = e^{(-2\pi i)/N}$, and *N* is the number of calculation points of the Fourier transform. The calculation of the signal spectrum is the signal's mean square amplitude of Fourier transformation, that is, a power spectrum, defined as:

$$P_{ss} = X(k)^2/T \qquad (3)$$

where $P_{ss}$ is the power spectrum estimation, $X(k)$ is the result of signal $x(n)$ after Fourier transform and *T* is the signal length. The power spectrum density function of each point can be calculated. One can draw a double logarithmic plot of the function (log*P* - log$\omega$). The log*P* is spatial frequency and log$\omega$ is power spectrum density. An approximately nonlinear fitting relationship can be written as:

$$G(\omega) = G(\omega_0)(\omega/\omega_0)^{-\eta} \qquad (4)$$

where $G(\omega)$ is signal power spectrum density, $\omega$ is spatial frequency, $\omega_0$ is reference spatial frequency. $G(\omega_0)$ is signal transform coefficient, the value of the signal power spectrum at $\omega_0$, and $\eta$ is the frequency index, representing the frequency structure of the power spectrum density.

After logarithmic change of Eq. 1 and Eq. 4, then fitting Eq. 4 by a linear regression, the coefficient of the regression fitting is $\eta$. So the fractal dimension is defined as:

$$D = b/a - \eta \tag{5}$$

Fig. 1 shows the logarithmic graph of neutron and gamma spectra. The single neutrons are generated using the method in Ref. [18]. It is found that neutrons and gamma-rays have different regression coefficients and spectral characteristics. By means of this difference one can distinguish neutrons from gamma-rays.

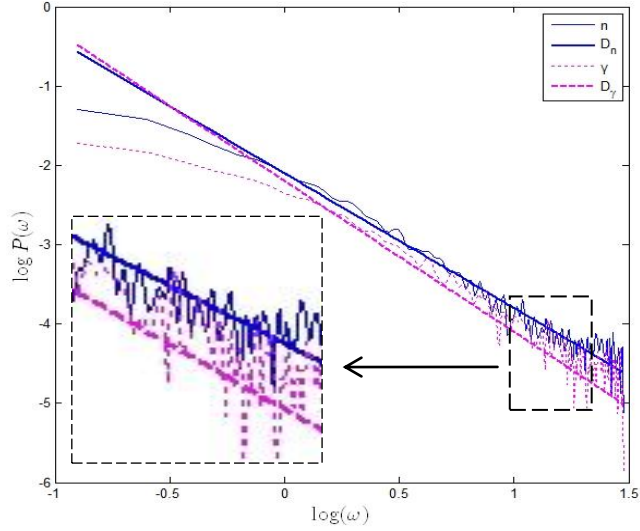

Fig.1 Llogarithmic graph of neutron and gamma spectra.

## III. CALCULATION ON RESULTS AND DISCUSSION

1. Comprehensive evaluation method of discrimination effects

Figure of Merit (FoM) and Discrimination Error Ratio (DER) have been used to evaluate discriminant effects [20], respectively. FoM is calculated using the following equation

$$FoM = \frac{S}{FWHM_\gamma + FWHM_n} \tag{6}$$

where $S$ is the distance between neutron peak and gamma-ray peak, $FWHM_\gamma$ is for gamma-ray peaks, and $FWHM_n$ is for neutron peaks. The larger the FoM the better the resolution. In most papers, only FoM is used as an evaluation criterion. However, FoM can only demonstrate the degree of separation in discrimination of neutrons and gamma-rays; it cannot show the purity of the discrimination. Accordingly we consider the second method to give a comprehensive evaluation.

The DER is defined as the ratio of the number of wrongly discriminated particles and the total number of all kinds of particles. $DER_\gamma$, $DER_n$ and $DER_{total}$ can be calculated using the following equations.

$$\begin{cases} DER_\gamma = \left|\dfrac{N_\gamma - N_\gamma^*}{N_\gamma}\right| \times 100\% \\ DER_n = \left|\dfrac{N_n - N_n^*}{N_n}\right| \times 100\% \\ DER_{total} = \left|\dfrac{(N_\gamma - N_\gamma^*) + (N_n - N_n^*)}{N_\gamma + N_n}\right| \times 100\% = \dfrac{N_\gamma}{N_\gamma + N_n} DER_\gamma + \dfrac{N_n}{N_\gamma + N_n} DER_n \end{cases} \quad (7)$$

where $N_\gamma$ is the total number of gammas, and $N_n$ is the total number of neutron pulses. $N_\gamma^*$ is the gamma count value discriminated by different methods such as the digital charge integration method, pulse gradient method, and fractal spectrum method. $N_n^*$ is the neutron pulse count value discriminated by the same three methods. Using this kind of criterion for evaluation, a smaller DER indicates better discrimination ability.

Note that FoM has the advantage that it can be calculated from experimental data; however, it is not the best parameter to determine which algorithm performs best. The DER has the advantage that it can be used to evaluate an algorithm in a simulation. Therefore, both FoM and DER are used in this paper to evaluate the proposed fractal spectrum performance. In addition, we note that simulated data are used in this paper and are based on a specific detector EJ-301 scintillator [20]. The detector parameters used are the same as those in Ref. [20].

## 2. Fractal spectrum for n/γ discrimination

A fractal spectrum was used to deal with 5000 mixed neutron and gamma signals which were generated by simulations in Ref. [20] (the noise parameter $\delta = 0.03$). The result is shown in Fig. 2. It can be seen that the resolution is very good and the value of FoM is 2.19.

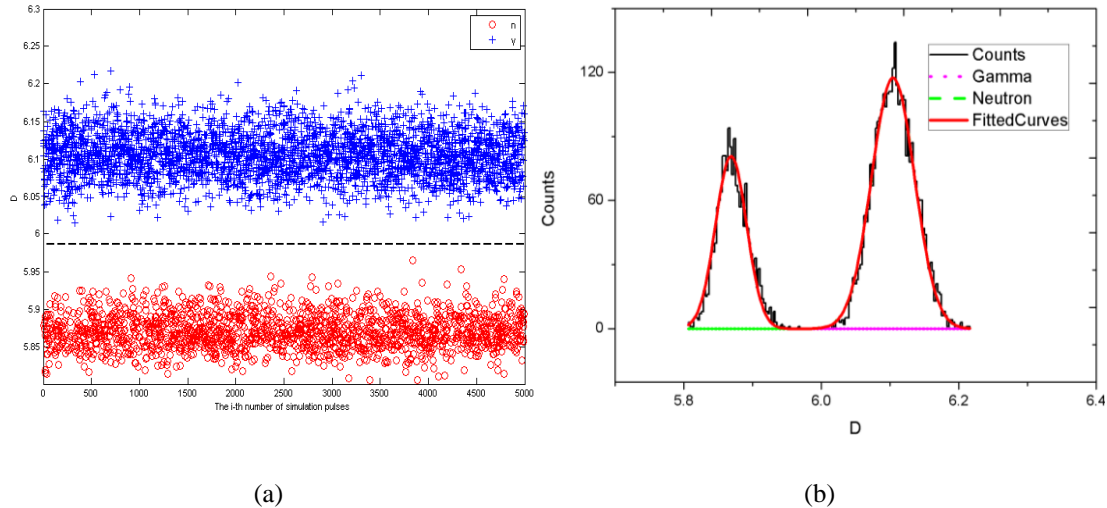

(a)  (b)

Fig.2 Discrimination results of 5000 mixed *n/γ* signals by fractal spectrum

(a) *n/γ* discrimination results, $\delta = 0.03$; (b) *n/γ* discrimination with FoM=2.19.

## 3. Impact of high frequency noise

The impact of three kinds of noise signals ($\delta = 0.01, 0.05, 0.10$) on discrimination ability is investigated in this subsection. A larger value of $\delta$ corresponds to higher noise. The overlap parameter τ is arbitrarily assumed to range from 20 to 50.

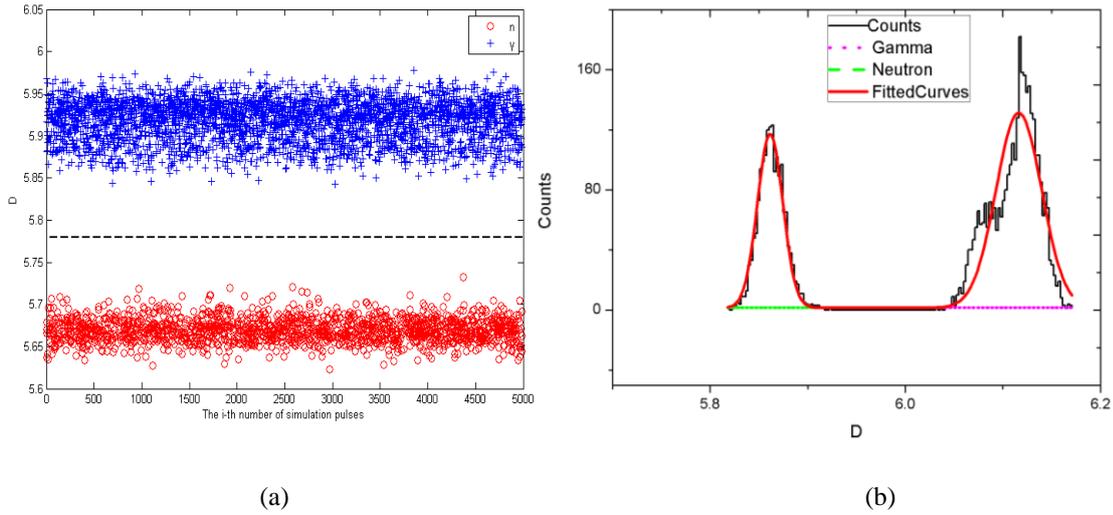

(a) (b)

Fig.3 Discrimination results of 5000 mixed n/γ signals by fractal spectrum.
(a) *n/γ* discrimination results; (b) *n/γ* discrimination with *σ* = 0.01, FoM=3.52.

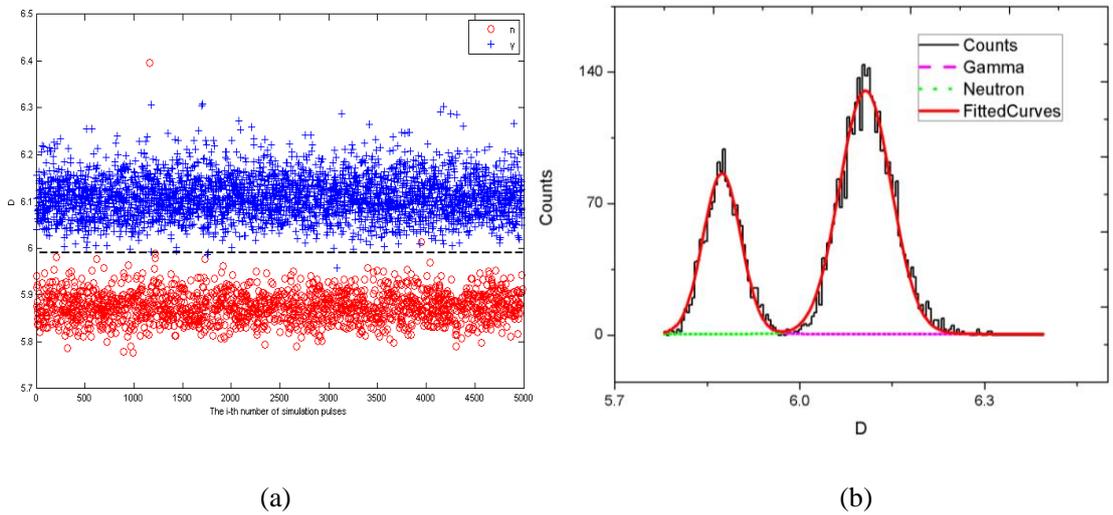

(a) (b)

Fig.4 Discrimination results of 5000 mixed n/γ signals by fractal spectrum.
(a) *n/γ* discrimination results; (b) *n/γ* discrimination with *δ* = 0.05, FoM = 1.61.

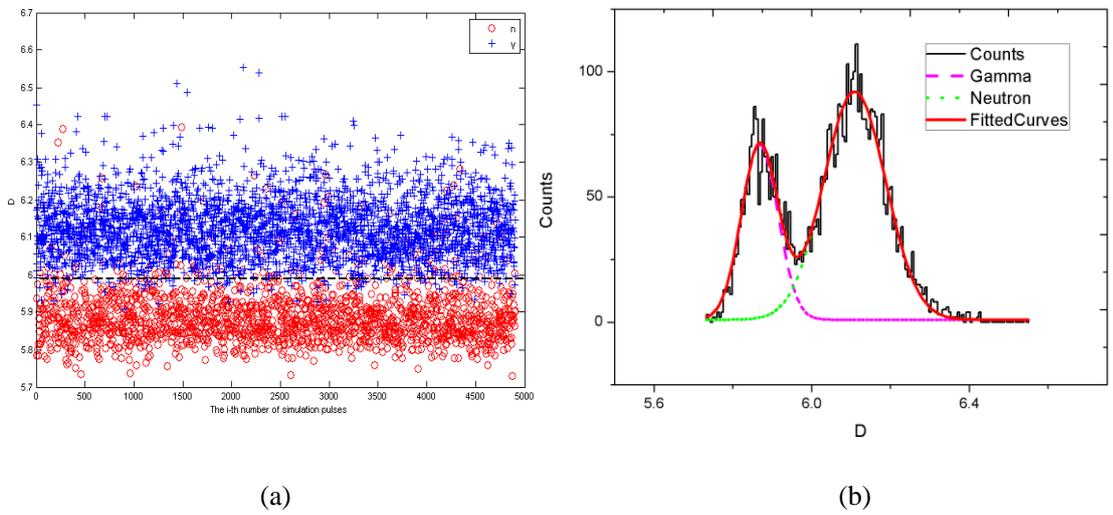

(a) (b)

Fig.5 Discrimination results of 5000 mixed n/γ signals by fractal spectrum.
(a) *n/γ* discrimination results; (b) *n/γ* discrimination with *δ* = 0.1, FoM = 0.96.

Comparing Fig. 3 to Fig. 5, it can be seen that with the increase of the noise signal, the resolution boundary becomes increasingly blurred, the corresponding FoM values decrease, and DER values increase (see Table 1).

TABLE 1. DER values of fractal spectrum with different kinds of noise

|  | $\delta = 0.01$ | $\delta = 0.03$ | $\delta = 0.05$ | $\delta = 0.10$ |
| --- | --- | --- | --- | --- |
| γ ray | 0.00 | 0.00 | 3.86 | 5.10 |
| Neutron | 0.00 | 0.00 | 2.44 | 3.64 |
| **Total** | **0.00** | **0.00** | **3.15** | **4.37** |

## 4. Impact of pulse pile-up

Keeping the noise standard deviation constant, with $\delta = 0.03$, we choose three different overlap conditions ($\tau_1$= 60~100$ns$, $\tau_2$= 120~160$ns$, $\tau_3$= 160~200$ns$), to study how overlapping peaks affect the fractal spectrum method.

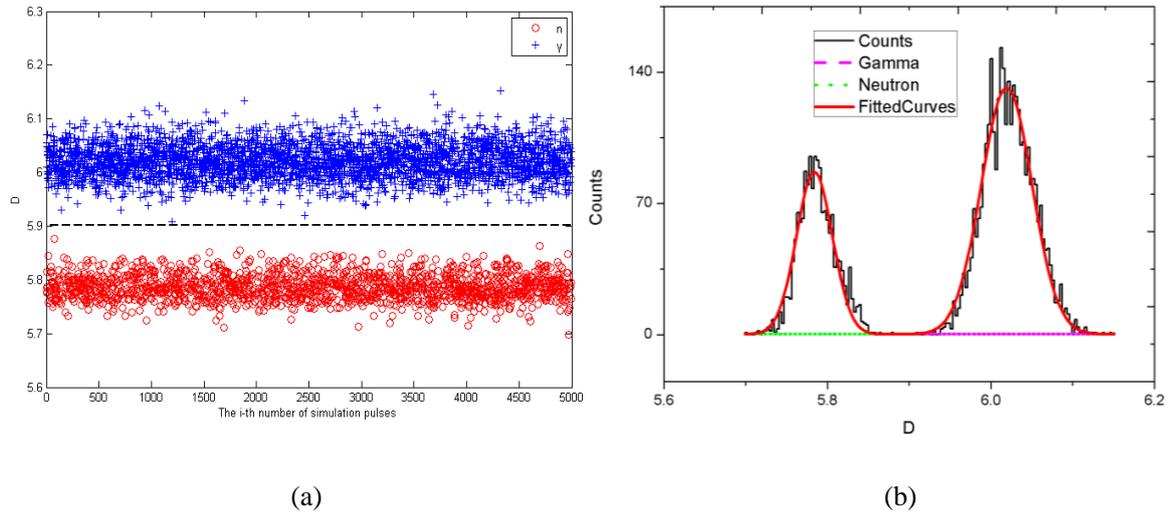

(a)          (b)

Fig.6 Discrimination results of 5000 mixed n/γ signals by fractal spectrum.

(a) *n/γ* discrimination results; (b) *n/γ* discrimination with 60 ~ 100$ns$, FoM=2.21.

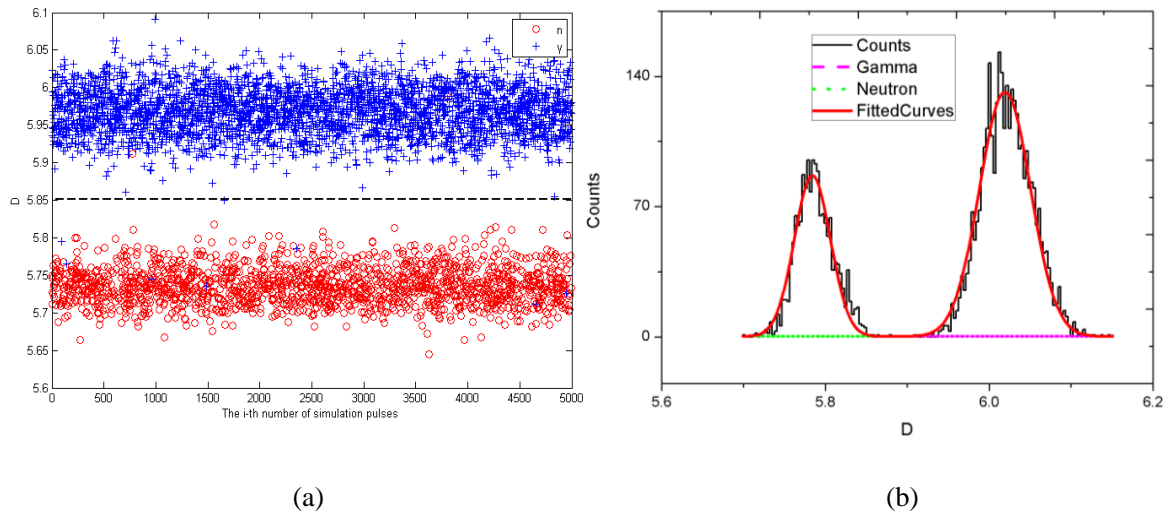

(a)          (b)

Fig.7 Discrimination results of 5000 mixed n/γ signals by fractal spectrum.

(a) *n/γ* discrimination results; (b) *n/γ* discrimination with 120 ~ 160*ns*, FoM=2.08.

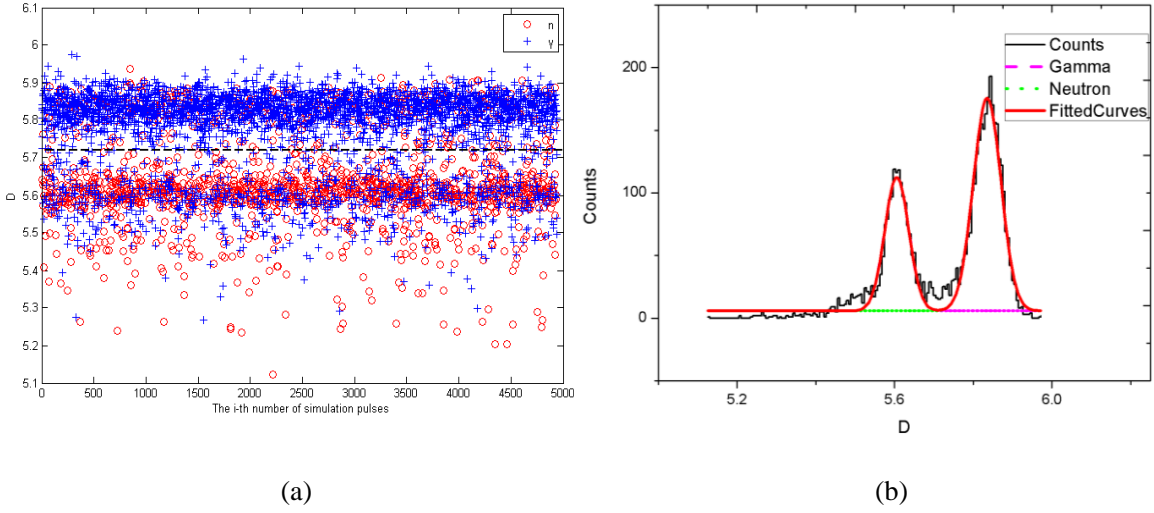

(a)　　　　　　　　　　　　　　　(b)

Fig.8 Discrimination results of 5000 mixed n/γ signals by fractal spectrum.

(a) *n/γ* discrimination results; (b) *n/γ* discrimination with 160 ~ 200*ns*, FoM=1.75.

One can observe that more and more neutrons are mistaken for gamma-rays with the increase of the value of *τ* from Fig. 6(a) to Fig. 8(a). That is, the probability of identifying a gamma-ray as a neutron gradually becomes larger. However, the value of the FoM becomes smaller at first, and then larger. It means that FoM does not satisfactorily describe resolution accuracy. Therefore, it is necessary to check the DER in the discrimination effect. It can be seen from Table 2 that DER increases with increasing *τ*.

TABLE 2. DER value of fractal spectrum under pulse pile-up

|  | τ=60 ~ 100 | τ=120~160 | τ=160~200 |
|---|---|---|---|
| γ ray | 0.00 | 0.31 | 28.07 |
| Neutron | 0.00 | 0.15 | 13.13 |
| **Total** | **0.00** | **0.23** | **20.60** |

5. Comparisons and discussion

1) Influence of high frequency noise

We arbitrarily set different values of *δ* to regulate different signal noise, and then studied the impact of high frequency noise on the three methods, namely, fractal spectrum, digital charge integration and pulse gradient. We note that the selection of parameters in the three methods is of vital importance. Different parameter settings may lead to different discriminating performance. Therefore, the optimal parameters in the three methods should be carefully selected in order to fairly compare the discrimination effects among the three methods. In the digital charge integration method, we use the complete integral and fast component approach. The integration times for the complete and fast component were 140 ns and 30 ns, respectively. In the pulse gradient method, values ranging within [15, 25] ns are believed to be optimal. Based on our studies, the value was set to be 22 ns and the threshold of the pulse height was set to be 0.5. In the fractal spectrum method, the threshold of pulse height was also 0.5. The sampling frequency was set to be 60, and the sampling points set to be 480. The values for the three methods were calculated, and the statistical results are shown in Table 3.

TABLE 3. FoM values under different high frequency noise signals

|  | Digital charge integration | Pulse gradient | Fractal spectrum |
|---|---|---|---|
| $\delta$ =0.01 | 5.27 | 3.66 | **3.52** |
| $\delta$ =0.03 | 1.81 | 1.34 | **2.19** |
| $\delta$ =0.05 | 1.13 | 0.74 | **1.61** |
| $\delta$ =0.10 | 0.57 | 0.32 | **0.96** |

With the increase of $\delta$, one can see that the values of FoM in the three methods decrease, indicating that high frequency noise has an impact on all three methods. For low frequency noise, the value of FoM in the digital charge integration method is the largest. For high frequency, the value of FoM in the pulse gradient method is minimal, while that for the fractal spectrum is relatively high, about two times that of the digital charge integration method, and three times that of the pulse gradient method, indicating that the fractal spectrum is least sensitive to noise.

Table 4 shows overall discrimination error ratio for the three methods. As can be seen from the table, as the noise becomes larger, DER increases. The pulse gradient method is more sensitive to noise; when the $\delta$ value is at 0.03, the digital charge integration method and fractal spectrum are zero, while the pulse gradient method is 2.11. Comparing the three methods, the fractal spectrum method has better anti-noise ability than the other methods because it transformed data into the frequency domain space.

TABLE 4. Overall discrimination error ratio with different noise

|  | Digital charge integration | Pulse gradient | Fractal spectrum |
|---|---|---|---|
| $\delta$ =0.01 | 0.00 | 0.00 | **0.00** |
| $\delta$ =0.03 | 0.00 | 2.11 | **0.00** |
| $\delta$ =0.05 | 2.34 | 3.71 | **3.15** |
| $\delta$ =0.10 | 5.14 | 4.63 | **4.37** |

2) Impact of pulse pile-up on the three methods

Different values of $\tau$ were used to study the impact of pulse pile-up on the three methods, and the values of FoM shown in Table 5 were obtained.

TABLE 5. FoM values under different pile-up signals for the three methods

|  | Digital charge integration | Pulse gradient | Fractal spectrum |
|---|---|---|---|
| $\tau = 60 \sim 100$ | 2.22 | 1.15 | **2.21** |
| $\tau = 120 \sim 160$ | 2.16 | 1.10 | **2.08** |
| $\tau = 160 \sim 200$ | 1.14 | 0.84 | **1.75** |

With τ increasing, pulse pile-up becomes more and more serious. The FoM of the three methods shows a decreasing tendency, but FoM as an evaluation criterion does not well reflect the correct rate of separation. According to the DER standard, as shown in Table 6, with increasing τ, the values of DER grow larger. Comparing the three methods, the pulsed gradient method is more sensitive to overlapping peaks, while the fractal spectrum method demonstrates the best performance in a variety of overlapping cases.

TABLE 6. DER values with different pile-up signals

|  | Digital charge integration | Pulse gradient | Fractal spectrum |
|---|---|---|---|
| $\tau = 60 \sim 100$ | 0.00 | 5.38 | **0.00** |
| $\tau = 120 \sim 160$ | 0.42 | 18.11 | **0.23** |
| $\tau = 160 \sim 200$ | 21.00 | 30.00 | **20.60** |

## IV. CONCLUSION

In a mixed neutron and gamma radiation field, the implementation of rapid and accurate discrimination of the particles is particularly important. Random mixed neutron and gamma signals in a liquid scintillator detector with different noise and different pulse pile-ups were simulated. Figure of merit and average discriminant error ratio were used together to evaluate the discrimination effects. The figure of merit reflects the degree of separation in particle discrimination, and average discriminant error ratio reflects the accuracy of discrimination.

This paper introduced an n/γ fractal spectrum discriminant approach based on different pulse shape characteristics induced by neutrons and gamma-rays. The proposed approach was compared with the digital charge integration method and pulse gradient method. The fractal approach exhibits the best discriminant performance among three methods, followed by the digital charge integration method, with the pulse gradient method third. In addition, this paper also discussed the influence of high frequency noise and pulse pile-up on the three methods. The calculation results indicate that the fractal spectrum is the least sensitive to high frequency noise and pulse pile-ups. It means that this approach has the advantages of anti-noise and high discrimination ability. At present we are seeking real data of neutrons and gamma-rays from a mixed radiation field to validate the effectiveness of the proposed approach and will report this in future.